\documentclass[aps,twocolumn,a4paper]{revtex4}
\usepackage[colorlinks=true, pdfstartview=FitV, linkcolor=blue, citecolor=red, urlcolor=magenta, breaklinks=true]{hyperref}
\usepackage{graphicx}
\usepackage[all]{xy}
\usepackage{amsmath}
\usepackage{amssymb}
\usepackage{color}
\usepackage{subeqnarray}
\usepackage{subfigure}
\usepackage{epstopdf}

\newcommand{\bes}{\begin{subequations}}
\newcommand{\ees}{\end{subequations}}
\def\ben{\begin{eqnarray}}
\def\een{\end{eqnarray}}
\newcommand{\bens}{\begin{subeqnarray}}
\newcommand{\eens}{\end{subeqnarray}}
\def\be{\begin{equation}}
\def\ee{\end{equation}}

\def\cos{\text{cos}}

\def\sin{\text{sin}}
\def\sech{\text{sech}}

\def\sec{\text{sec}}
\def\ln{\text{ln}}

\begin{document}
\title{Analytical study of kinklike structures with polynomial tails}
\author{D. Bazeia$^a$\footnote{Corresponding author; email: dbazeia@gmail.com}, R. Menezes$^{a,b}$, and D.C. Moreira$^a$}
\affiliation{$^a$Departamento de F\'isica, Universidade Federal da Para\'iba, 58051-970, Jo\~ao Pessoa, PB, Brazil}
\affiliation{$^b$Departamento de Ci\^encias Exatas, Universidade Federal da Para\'\i ba, 58297-000 Rio Tinto, PB, Brazil}
\begin{abstract}
This work deals with models described by a single real scalar field in two-dimensional spacetime. The aim is to propose potentials that support massless minima and investigate the presence of kinklike structures that engender polynomial tails. The results unveil the presence of families of asymmetric solutions with energy density and linear stability that behave adequately, enhancing the importance of the analytical study. We stress that the novel topological structures which we find in this work engender long range interactions that are of current interest to statistical mechanics, dipolar quantum gases and the study of quantum information with Rydberg atoms.
\end{abstract}
\maketitle

\section{Introduction}

The study of  nonlinear phenomena is of great interest to Physics and many effects related to nonlinearity acquire relevant role in the description of spatially localized structures \cite{v,sut,vacha}.  In $(1,1)$ spacetime dimensions, in particular, there are scalar field models that support localized structures known as {\it  kinks}, which appear due to the set of degenerate minima that characterize the several topological sectors of the systems. Each topological sector provides a kink-antikink pair of solutions, which has been explored in several scenarios; see, e.g., Refs.~\cite{v,sut,vacha,GT,BD,vacha0,gui,lobo,brito,BLM,BMM,moreira1,moreira3,elisama} and references therein. 

In the potential that defines the model, a feature of a given minimum is the mass associated with it, which is relevant, in particular, to indicate how the scalar field behaves as it approaches the minimum asymptotically. As it is known, for masses different from zero, we find an exponential decay of the scalar field configuration to the minimum of the potential. This feature is interesting, for example, for the study of the force between kinks, where one usually obtains short-range interactions mediated by the asymptotic behavior of the kinklike solutions \cite{manton,PT}. In addition, the existence of topological sectors linking minima with non-vanishing masses may result in the presence of excited modes in the spectrum of the stability equation related to resonances arising in kink/anti-kink scattering process \cite{takyi,gani,oliveira,riazi,gmo,gani5,dorey,kar,gani2,adalto1,adalto2}. 

Families of models with scalar field solutions with distinct qualitative aspects may emerge when one considers potentials with minima presenting zero-mass behavior. The asymptotic behavior of the field solution approaching these minima may obey a power law, instead of the above mentioned exponential decay. Thus, the kink has an extended tail that simulates long range interaction since it has a long range force emerging from it \cite{gmo,riazi}. This type of model may appear when one increases the degree of the polynomial potential \cite{lohe,khare}, or under the presence of non-polynomial interactions.

Structures that engender long range interactions are important because they can be used in applications in a diversity of scenarios of practical interest, for instance, in the statistical mechanics of solvable models with long range interactions \cite{SM}, in the investigation of dipolar quantum gases \cite{QG} and in the study of quantum information with Rydberg atoms \cite{QI}. In these cases, the long range interactions are in general mediated by forces that include the dipole-dipole and the van der Waals type.

Despite the natural interest in exploring possible new effects arising from topological structures with polynomial tails, there are still few analytical models with these properties \cite{gmo,riazi}, which are briefly reviewed in Sec. II, where we also describe the formalism to be used in the current work. We then go further and describe new models, with the potentials having massless minima that support analytical kinklike solutions with distinct polynomial asymptotic behavior. We search for asymmetric solutions, in a way similar to the case already investigated in \cite{moreira2,marques,roldao,melfo}, but when one has at least one of the minima with zero mass, other relevant characteristics are found. The families of models to be studied present analytical solutions and have two distinct types of asymmetry. They are investigated in Sec.~III, and the first family supports solutions that connect two different types of minima, one which is massless and the other with a nonzero mass. The second family of models engenders solutions that connect massless minima, but now they have different dominant terms in the corresponding asymptotic expansions. We finish the work in Sec.~IV, briefly reviewing the main results and adding comments on new issues related to the current investigation. 

\section{Generalities}

In the case of a single real scalar field, the Lagrangian density can be written as
\be\label{ori}
\mathcal{L}(\phi,\partial_{\mu}\phi)=\frac{1}{2}\partial_{\mu}\phi\partial^{\mu}\phi-V(\phi),
\ee
where $ V(\phi) $ is the potential used to describe the behavior of the scalar field, identified by $\phi $. The equation of motion has the form
 \begin{equation}\label{boxeq}
\partial_{\mu}\partial^{\mu}\phi+\frac{dV}{d\phi}=0.
\end{equation}
Moreover, in $(1,1)$ dimensions we have $\phi=\phi(x,t)$ and thus equation \eqref{boxeq} represents a  second order differential equation which, for static configurations, becomes $\phi''(x)={dV}/{d\phi},$
where prime denotes derivative with respect to the spatial coordinate. Here we are using the metric $\eta_{\mu\nu}=\text{diag}(1,-1)$. When the potential has a set of degenerate minima, one can define topological sectors linking each pair of consecutive minima.  Appropriate boundary conditions on the scalar field imply that its solution in the coordinate space must have the shape of a {\it kink}. The kink is a function with the form similar to that of the hyperbolic tangent and is characterized by the existence of a topological current, usually defined as $j^{\mu}=\epsilon^{\mu\nu}\partial_{\nu}\phi$, where $\epsilon^{\mu\nu}$ is the antisymmetric symbol in two dimensions with $\epsilon^{01}=1$. This leads us with the quantity  $Q=\phi (\infty)-\phi (-\infty)$, which is the {\it topological charge} of the kink. The use of appropriate boundary conditions requires that we choose $\phi(\infty)$ and $\phi(-\infty)$ as two distinct and consecutive minima of the potential, and this imposes the presence of at least two degenerate minima in the potential. We can add a Lorentz boost to the solution to make it a traveling wave. This means that if $\phi(x)$ is a solution, so is $\phi(y)$, for $y=\gamma\,(x-vt)$, with $\gamma=\sqrt{1-v^2}$. This is important to investigate scattering of kinks and anti-kinks, a subject currently under consideration, to be reported elsewhere \cite{bbg}.

The energy-momentum tensor associated to the model is $T^{\mu\nu}=\partial^{\mu}\phi\partial^{\nu}\phi-\eta^{\mu\nu}\mathcal{L}$. It leads us to the energy density which is giving by 
\begin{equation}
\label{rho1}
T^{00}=\rho (x)=\frac{1}{2}\phi'^2+V(\phi).
\end{equation} 
Here we consider non-negative potentials. This allows that we use an auxiliary function, $W=W(\phi)$, and write the potential of the model as
$V(\phi)=(1/2)({dW}/{d\phi})^2.$ In this case, the total energy of the solutions can be minimized if the field configurations obey the simpler first order differential equations $\phi'=\pm{dW}/{d\phi},$ with the two signs giving rise to kink and anti-kink. In general, kinks are described by smooth and increasing functions, and anti-kinks by smooth and decreasing functions. A direct consequence of the use of $W$ is that now the energy density becomes $\rho(x)=\phi'^2=(dW/d\phi)^2$. The procedure has the advantage that once one knows the function $ W(\phi) $, we directly find the minimal energy of the solution, which can be written as the absolute value of $\Delta W$, for $\Delta W= W(\phi(\infty))-W(\phi(-\infty))$.
This is important because we can obtain the total energy of the solution without knowing the solution explicitly. 

To analyze the linear stability of the static solutions, one takes
$\phi(x,t)=\phi(x)+\eta(x,t)$; for small fluctuations we write $\eta(x,t)=\sum_k \eta_k(x)\cos (\omega_k t)$ and use the equation of motion \eqref{boxeq} to find the linear equation 
\begin{equation}\label{stabilityequation}
\left(-\frac{d^2}{dx^2}+v(x)\right)\eta_k(x)=\omega_k^2\,\eta_k(x).
\end{equation}
This is a Schr\"odinger-like equation with a set of eigenfunctions $\left\{\eta_k(x)\right\}$, their respective energies $\left\{\omega_k^2\right\}$ and the potential $v(x)={d^2V}/{d\phi^2}{|}_{\phi=\phi(x)},$
which represents the stability potential of the system. Proving that equation \eqref{stabilityequation} does not have bound states with negative energy ensures stability of the model. The proof can be achieved as follows: for $V=(1/2)(dW/d\phi)$ , the Hamiltonian $\hat{H}=\left(-d^2/dx^2+v(x)\right)$ of the above equation \eqref{stabilityequation} can be decomposed in the form $\hat{H}=S^{\dagger} S$, with $S=d/dx-d^2 W/d\phi^2$. It implies that  $\hat{H}$ is non-negative and so we have $\omega_k^2\geq 0$ for all values of $k$. Moreover, the translation invariance of the models implies the existence of the zero mode, given by the solution of the equation $S\eta_0(x)=0$, which can be written as $\eta_0(x)=\phi'(x)$.

In order to understand how the tail of the kink behaves, we recall that the topological solution $\phi(x)$ has to go asymptotically to a minimum of the potential, which we generically denote by $\bar\phi$. We then perform a perturbation in the scalar potential using $\phi=\bar \phi+\delta$ in order to write
\begin{equation}\label{exp}
\frac{dV}{d\phi}\biggr|_{\bar\phi+\delta}= {\sum\limits_{j=1}^{}}\;a_j\;\delta^j ,
\end{equation}
where 
\begin{equation}
a_j=\frac{1}{j!}\frac{d^{(j+1)}V}{d\phi^{(j+1)}}\biggl|_{\bar\phi},
\end{equation}
identifies the coefficients of the above expansion. This implies that for $\delta=\delta(x)$ very small, it has to obey the equation
\begin{equation}
\delta^{\,\prime\prime}=a_k \delta^k,
\end{equation}
where $k$ is the lowest value of $j$ that gives a non vanishing coefficient $a_k$.  

If there is a (classical) mass associated to the minimum $\bar\phi$, one has $a_1=m^2$, and in this case the asymptotic behavior of the kink presents an exponential suppression, of the form $\phi(x)\approx \bar\phi-\alpha e^{-mx}$, for very large values of $x$, with $\alpha$ being a real and positive constant; however, if the minimum is massless, the asymptotic behavior of the topological structure is polynomial, with the corresponding power depending on the first nonzero coefficient that appears in the expansion \eqref{exp}, for $j>1$.
In this case, one can find solutions with tails that present a power law behavior. This shows that the specific form of the potential of the model contributes to control the tail of the topological configuration.

The program to be studied has two distinct objectives, the first one appearing in the current work, where we focus mainly on the construction of models which support analytical solutions that represent stable and localized topological structures with asymptotic behavior of the power law type. The other objective is part of a separate and longer investigation \cite{bbg}, with focus on the numerical simulations involving kink/anti-kink scattering processes that use the analytical results of the current work as structures to be considered in collisions, to see how the long range forces contribute to modify the standard scenario, where topological structures with exponential tails are used to scatter one another.

Although there are some works in the literature that explore models with properties similar to those discussed above, we noted that the subject still needs further investigation. In particular, the models 
\ben\label{modelexample}
V_1(\phi)&=&\frac{1}{2}\cos^4(\phi),\\
V_2(\phi)&=&\frac{1}{2}\left|1-\phi^2\right|^3,
\een
were studied before in \cite{riazi,gmo}, and have solutions given by
\bes\label{soluexample}\ben
\phi_1(x)&=&\pm\arctan (x) +k\pi, ~~~k\in \mathbb{Z},\\
\phi_2(x)&=&\pm\frac{x}{\sqrt{1+x^2}}.
\een\ees
These solutions approach their corresponding minima in a way proportional to $1/x$ and $1/x^2$, respectively. In this sense, 
in this work we add further results to the subject and, in the next section, we study two families of models, which support analytical solutions with distinct asymptotic behavior.

\section{New models}

Here we use natural units, with $\hbar=1=c$ and redefine the field and the space and time coordinates to work with dimensionless field and coordinates. We focus on models that engender topological structures that can be used to describe dipole-dipole, van der Waals and other long range interactions.

\subsection{Model I}

\begin{figure}
{\includegraphics[width=0.46\linewidth]{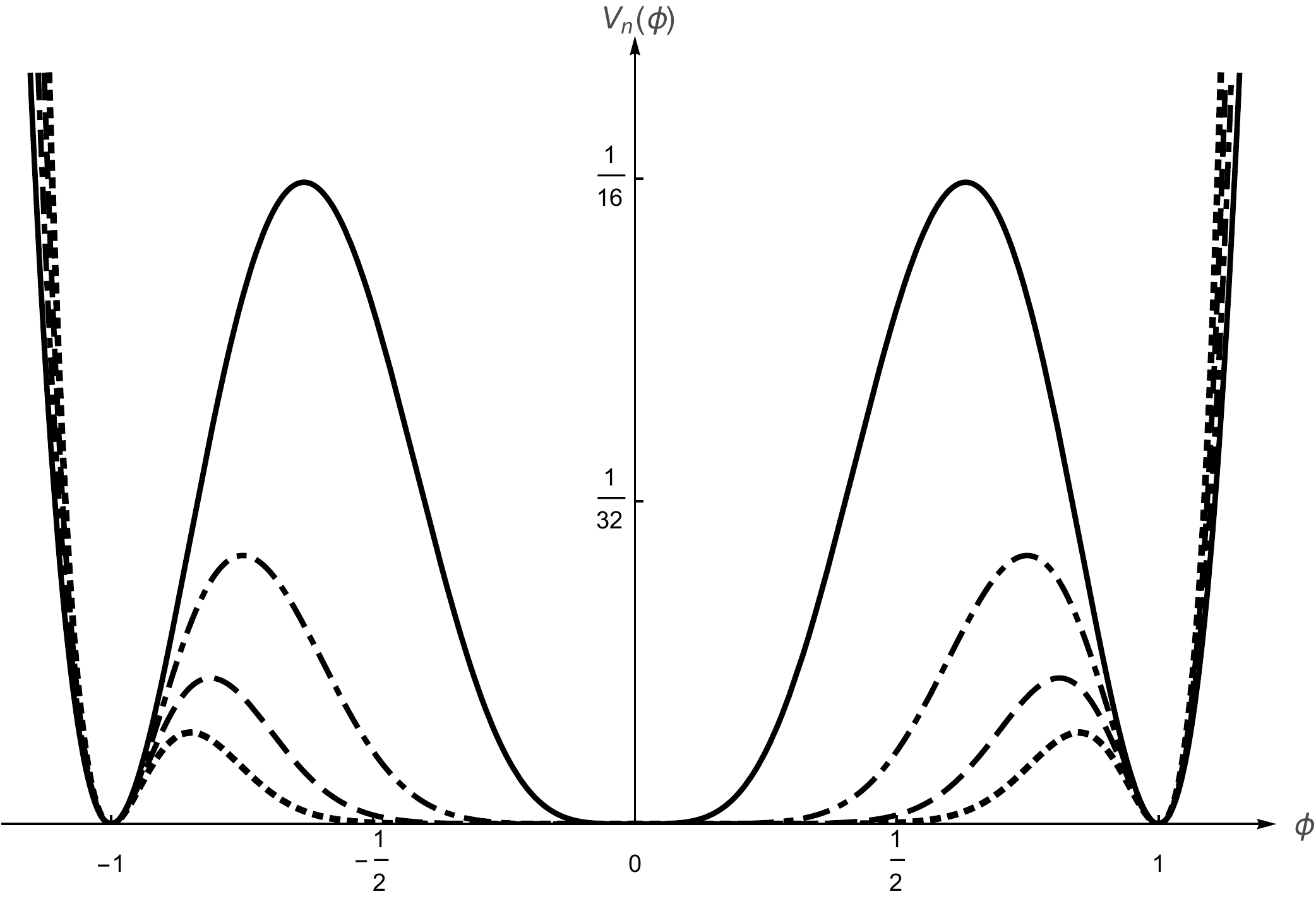}\label{seA}}
{\includegraphics[width=0.46\linewidth]{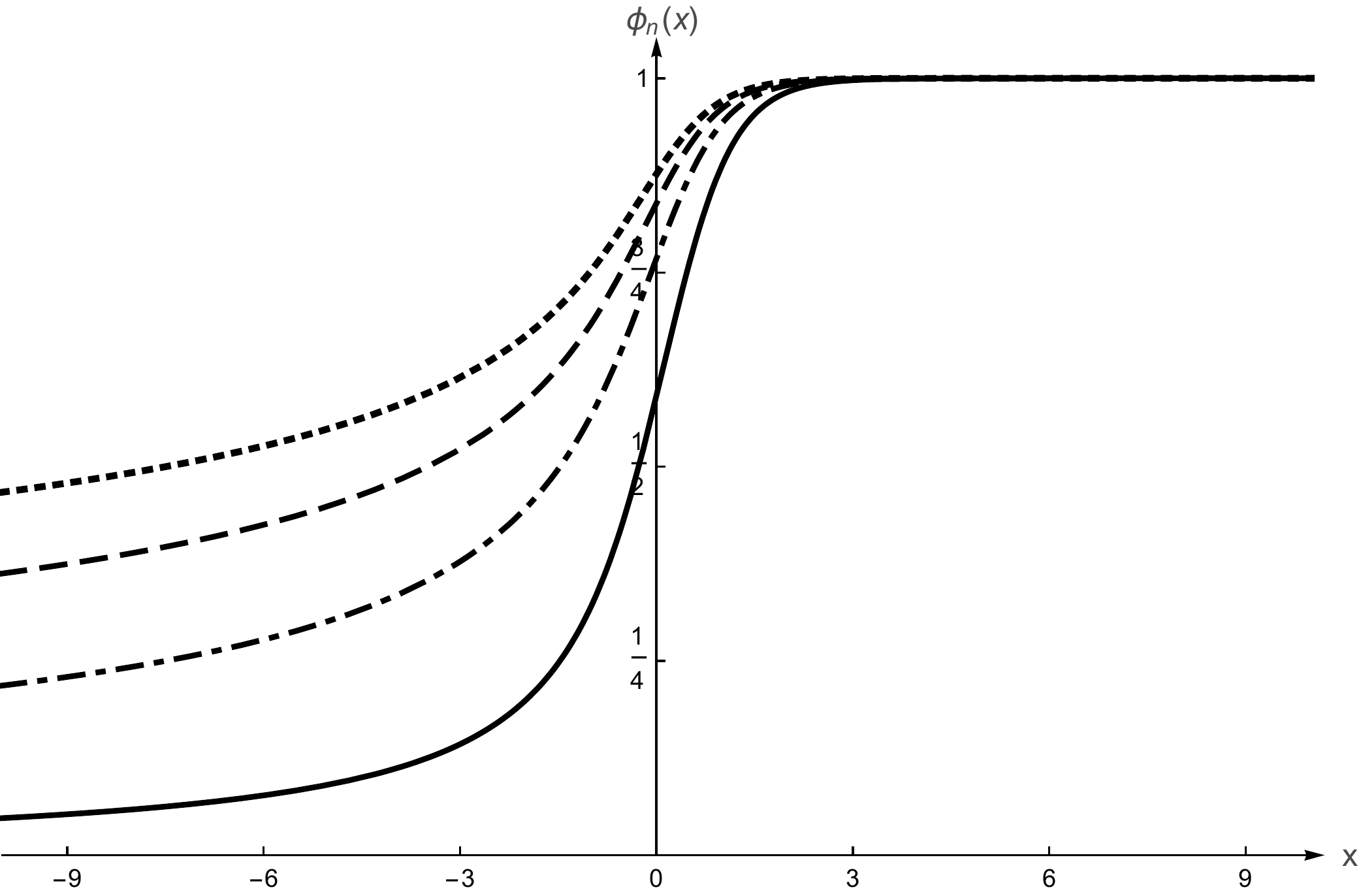}\label{seB}}
{\vspace{0.5cm}}
{\includegraphics[width=0.47\linewidth]{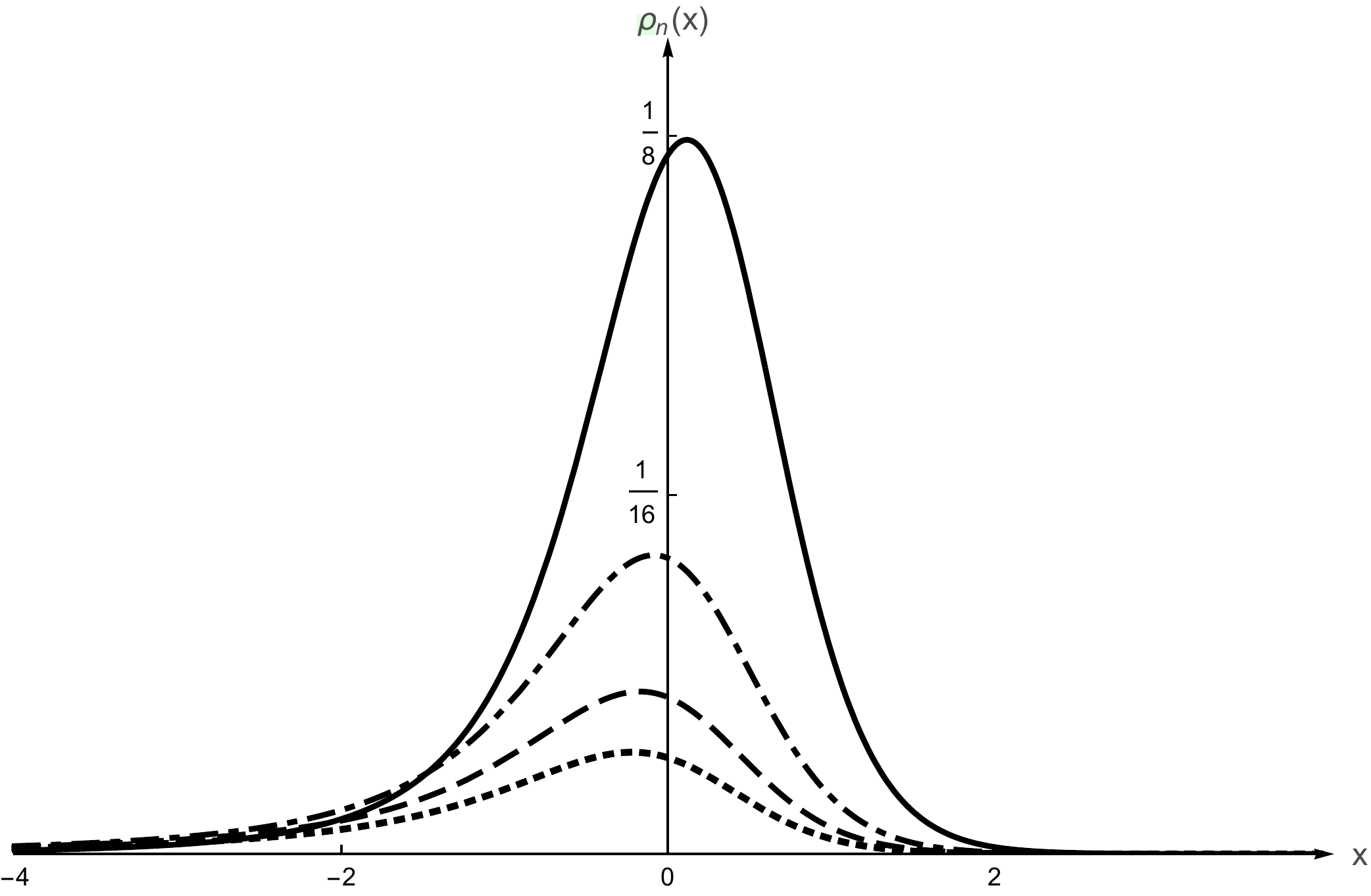}\label{seC}}
{\includegraphics[width=0.47\linewidth]{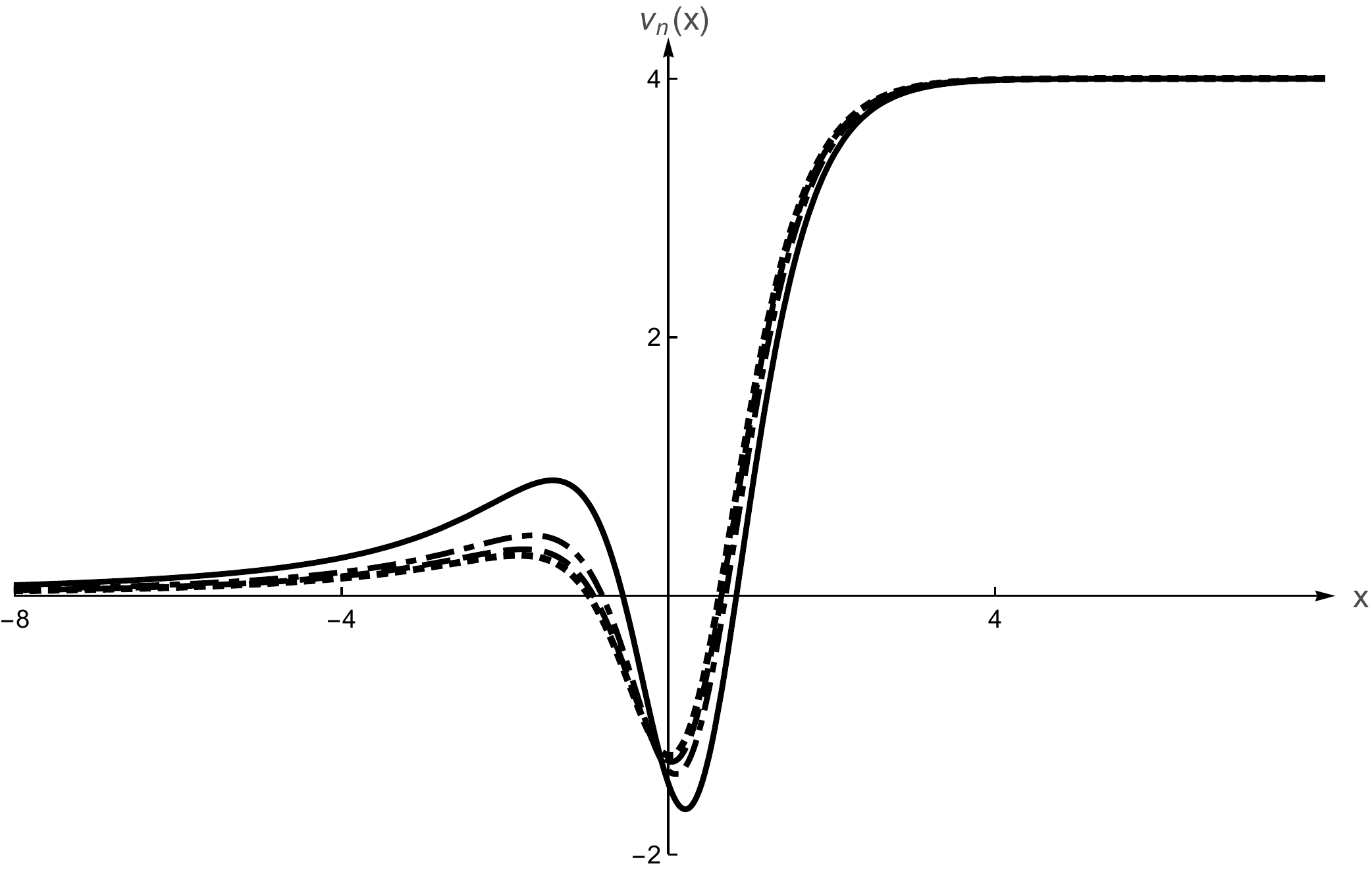}\label{seD}}
\caption{Illustration of the potential (top left), field solution (top right), energy density (bottom left) and stability potential (bottom right) for the model \eqref{pot1}. The solid, dot-dashed, dashed and dotted lines refer to $n=1$, $n=2$, $n=3$ and $n=4$, respectively.\label{fig1}}
\end{figure}

The first model to be studied is described by the potential
\begin{equation}\label{pot1}
V_n(\phi)=\frac{2}{n^2} \phi^{2 (n+1)} \left(1-e^{1-\left| \phi\right| ^{-n}}\right)^2,
\end{equation}
which is depicted in Fig.~\ref{fig1} for some values of $n=1,2,3,...\,$. The potential supports three degenerate minima, one at $\phi_0=0$ and two others, at $\phi_\pm=\pm1$. There are two topological sectors, but due to the presence of reflection symmetry we only need to study one of them, since the other can be obtained by symmetry operations. So we choose to deal with solutions in the right sector. 

When inserting the potential \eqref{pot1} into the equation of motion we find the second-order differential equation
\begin{equation}\label{2eommodel1}
\phi''\!=\!b_n e^{-2 \phi^{-n}}\!\!\left(\!e^{\phi^{-n}}\!\!\!- e\right)\!\left(\!\!\left(\!e^{\phi^{-n}}\!\!\!- e\right)\!\phi^n\!-\!\frac{en}{n\!\!+\!\!1}\!\right)\phi^{n\!+\!1}\!,
\end{equation}
where $b_n=4(n+1)/n^2$. Note that  in addition to the already identified minima we have two local maxima given by solutions of the transcendental equation
\begin{equation} 
e^{-1+\phi^{-n}}=1+\frac{n}{n+1}\phi^{-n}.
\end{equation}
This model admits the auxiliary function $ W(\phi)$, given  by
\begin{equation}\label{w1}
W\!(\phi)\!=\!c_n\left(\!\!\left(\!1-e^{1\!-\!\left|\phi\right|^{-n}}\!\right)\!|\phi|^{n\!+\!2}+e \Gamma\left(\!-\frac{2}{n},\left|\phi\right|^{-n}\!\right)\!\!\right)\!\!,
\end{equation}
where $c_n=\frac{2}{n (n+2)}$ and $\Gamma(a,b)$ is the incomplete Gamma function. Thus, the equation we need to solve is
\begin{equation}\label{1eommodel1}
\phi'=\frac{2 }{n}\left(1-e^{1-\phi^{-n}}\right) \phi^{n+1},
\end{equation}
which is satisfied by
\begin{equation}\label{sol1}
\phi_n(x)=\left(1+\ln \left(1+e^{-2 x}\right)\right)^{-1/n},
\end{equation}
when setting the condition $\phi_n(0)=\left(\ln(2e)\right)^{-1/n}$. This solution is asymmetric and connects the minima $\phi_0$ and $\phi_+$. Asymptotically, we have
$$\phi_n(x)\simeq\left\{
\begin{array}{c}
1-\frac{1}{n} e^{-2x}+\cdots, \text{if}~x\to\infty\\
{}\\
{}\left(2|x|\right)^{-1/n}+\cdots, \text{if}~x\to-\infty\\
\end{array}
\right.$$
The distinct asymptotic behavior comes from the mass associated to each one of the two minima, as it can be observed in Fig.~\ref{fig1}. At $\phi_+$ we have $m_+^2 =4 $, which lead us to an exponential decay as $x\to\infty$. At $\phi_0$ we have $m_0^2=0$, forcing the field to a power law depending of the model parameter as it approaches the symmetric minimum in the opposite limit. The existence of such behavior in the direction of the massless minimum indicates that the field approaches this regime in a slower way than that presented for the massive minimum. For the potential \eqref{pot1} at $\phi=\phi_0=0$ one finds that the $m-th$ derivative has the form ${d^m V}/{d\phi^m}|_{\phi=\phi_0}=m!$
for $m=2n+2$; all the lower order derivative vanish, so the greater the value of the parameter $n$, the higher the multiplicity of the zero at the origin, at the center of the potential \eqref{pot1}. 

This model is somehow similar to the standard $\phi^6$ model, which supports localized structure that decays asymmetrically, but engendering short range tails with exponential behavior. In the current model \eqref{pot1} the localized structure also decays asymmetrically, but now it has long range tail, and this will contributes to the scattering of kinks in a way different from the case of the $\phi^6$ model. This effect appears very clearly in the potential depicted in Fig.~\ref{fig1}, and it controls how the tail of the solution behaves asymptotically. 

The energy density of the kinklike solution is given by
\begin{equation}\label{ro1}
\rho_n(x)=\frac{4}{n^2}\frac{ \left(1+\ln \left(e^{-2 x}+1\right)\right)^{-\frac{2 (n+1)}{n}}}{ \left(e^{2 x}+1\right)^2}
\end{equation}
As expected, $\rho_n(x)$ reflects the asymmetry of the scalar field and is also depicted in Fig.~\ref{fig1}. When $x\to\pm\infty$ it behaves as follows
$$\rho_n(x)\simeq\left\{
\begin{array}{c}
\frac{4}{n^2} e^{-4x}+\cdots,\;\;\;\;\;\;~\text{if}~x\to\infty,\\
{}\\
{}\frac{4}{n^2}\left(2|x|\right)^{-\frac{2(n+1)}{n}}+\cdots,~\text{if}~x\to-\infty.\\
\end{array}
\right.$$ 
Integrating equation \eqref{ro1}, we find the total energy of the solution, which is given by
\begin{equation}
E_{n,BPS}=\frac{2e}{n(n+2)}\Gamma \left(-\frac{2}{n},1\right).
\end{equation}

The stability potential is 
\begin{equation}\label{sp1}
v_n(x)\!=\!-\sech^2 x\left(1\!-\!e^{2x}\!+\!\alpha_1\theta(x)\!+\!\alpha_2 e^{-2x}\theta^2(x)\right)\!\!,
\end{equation}
with $\alpha_1=(3+n)/n$, $\alpha_2=-(1+n)(1+2n)/n^2$ and we define the function $\theta(x)=\left(1+\ln\left(1+e^{-2x}\right)\right)^{-1}$. As depicted in Fig.~\ref{fig1}, the stability potential \eqref{sp1} has two distinct behaviors when $x\to\pm\infty$, resulting in a  mixed volcano/modified P\"oschl-Teller shape. When $x\to\infty$ the stability potential approaches $m_+^2$ and in the limit $x\to-\infty$ it approaches zero, which suggests that there is only one bound state, the zero mode, given by
\begin{equation}
\eta_0(x)=\frac{2}{n}\frac{\left(1+\ln\left(e^{-2 x}+1\right)\right)^{-\frac{n+1}{n}}}{e^{2 x}+1}.
\end{equation}

\begin{figure}[t]
{\includegraphics[width=0.46\linewidth]{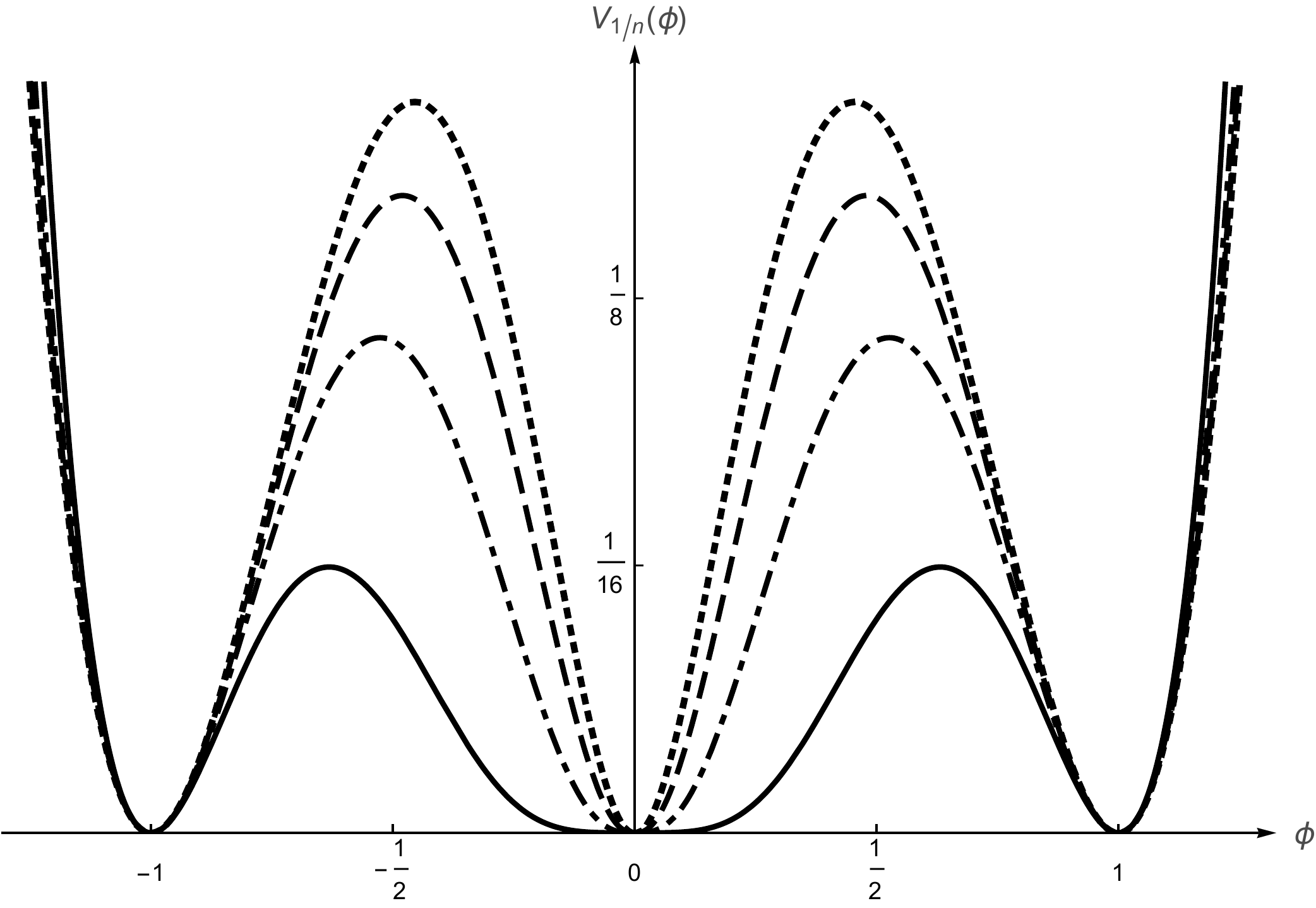}\label{seA}}
{ \includegraphics[width=0.46\linewidth]{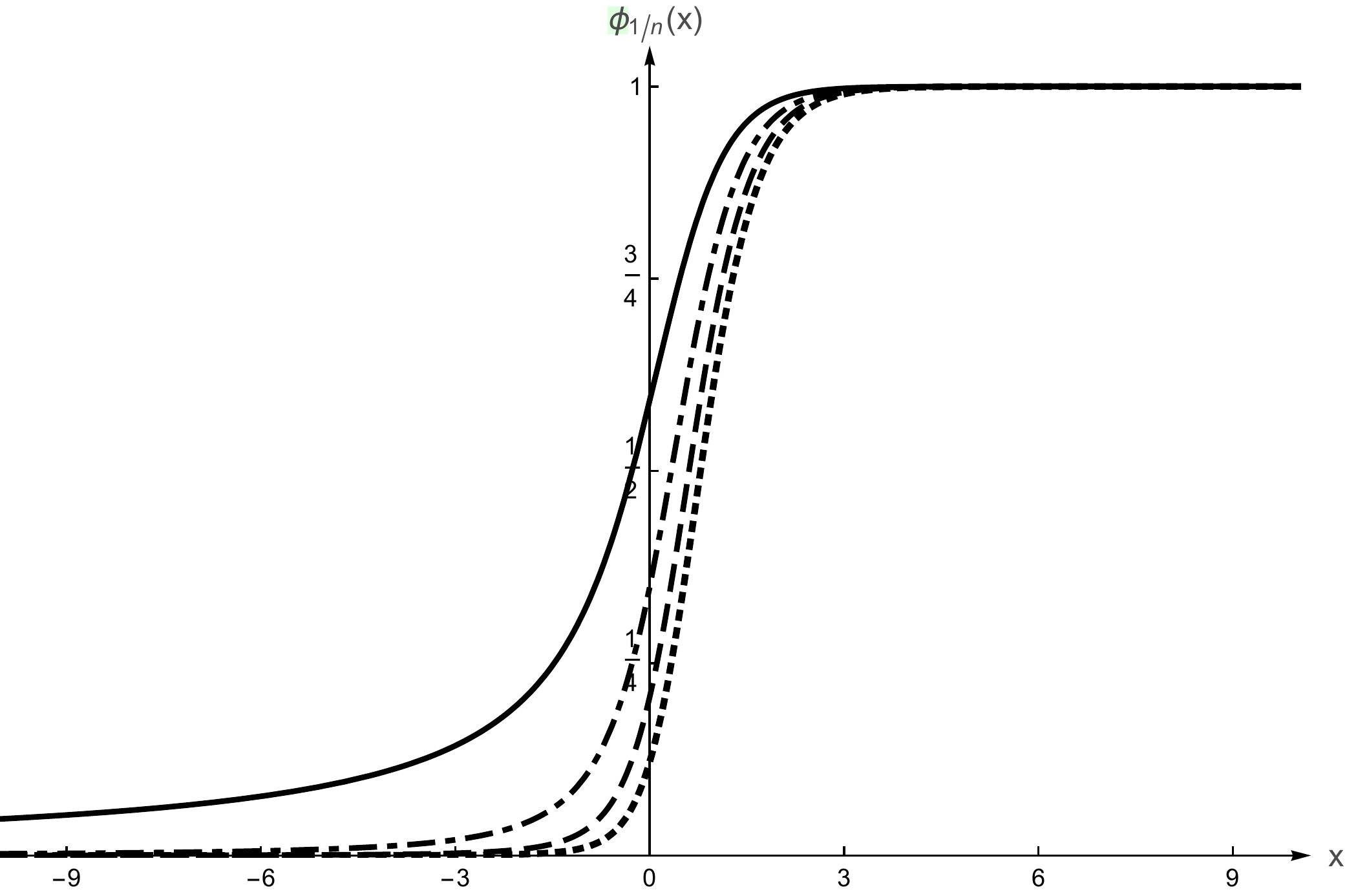}\label{seB}}
{\vspace{0.5cm}}
{\includegraphics[width=0.47\linewidth]{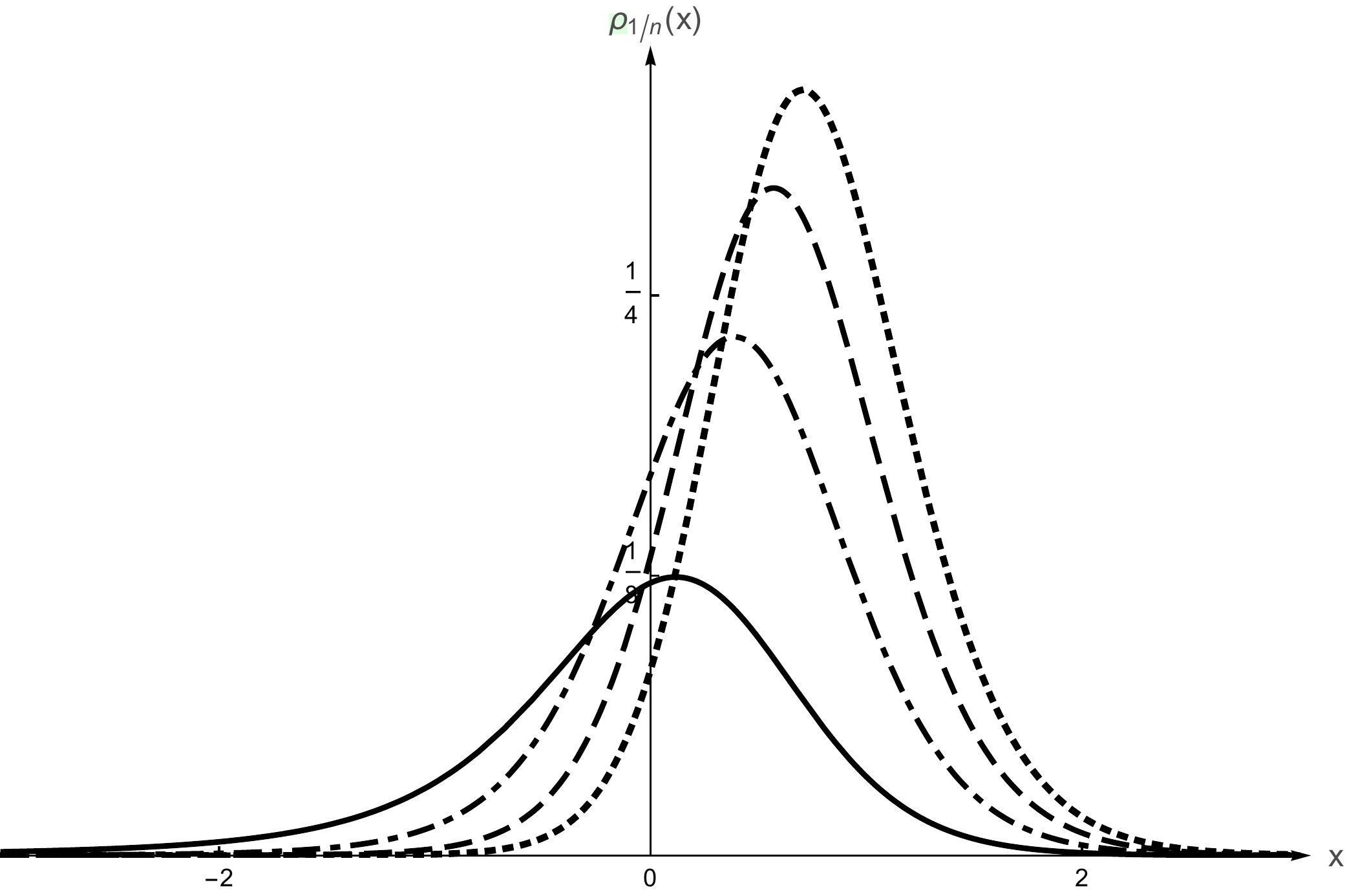}\label{seC}}
{\includegraphics[width=0.47\linewidth]{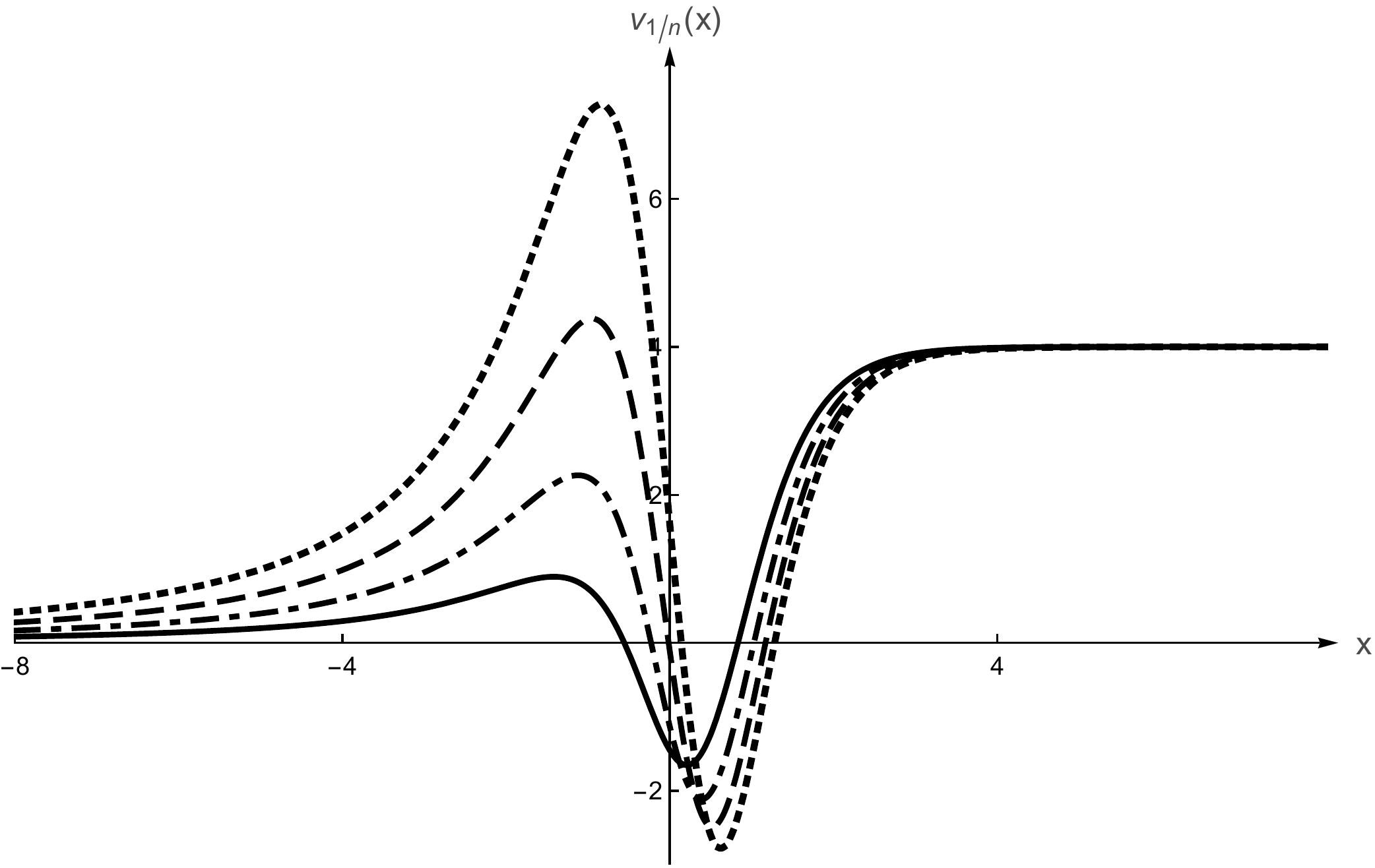}\label{seD}}
\caption{Illustration of the potential (top left), field solution (top right), energy density (bottom left) and stability potential (bottom right) for the model \eqref{pot1}, after changing $n\to1/n$. The solid, dot-dashed, dashed and dotted lines refer to $n=1$, $n=2$, $n=3$ and $n=4$, respectively. \label{fig2} }
\end{figure}

Note that the polynomial tail of \eqref{sol1} indicates a long-range interaction which depends of the specific choice of the parameter $n$. This shows that we can control the degree of the power law and, consequently, the force between kink and anti-kink. However, we can choose other ways of introducing the parameter in the system. For example, by performing the change $n\to 1/n $ in the potential \eqref{pot1}, we find a new model, which has another power law and implies a force between kinks with different degree.  This case is depicted in Fig.~\ref{fig2}. With the change $n\to 1/n$, the value of $n$ may now induce a dipole-dipole type of interaction, for $n=3$ or yet other powers, similar to the van der Waals force, so $n$ may be used to model long range interactions that occur in physical system of current interest \cite{SM,QG,QI}.

\subsection{Model II}

\begin{figure}[t]
{\includegraphics[width=0.46\linewidth]{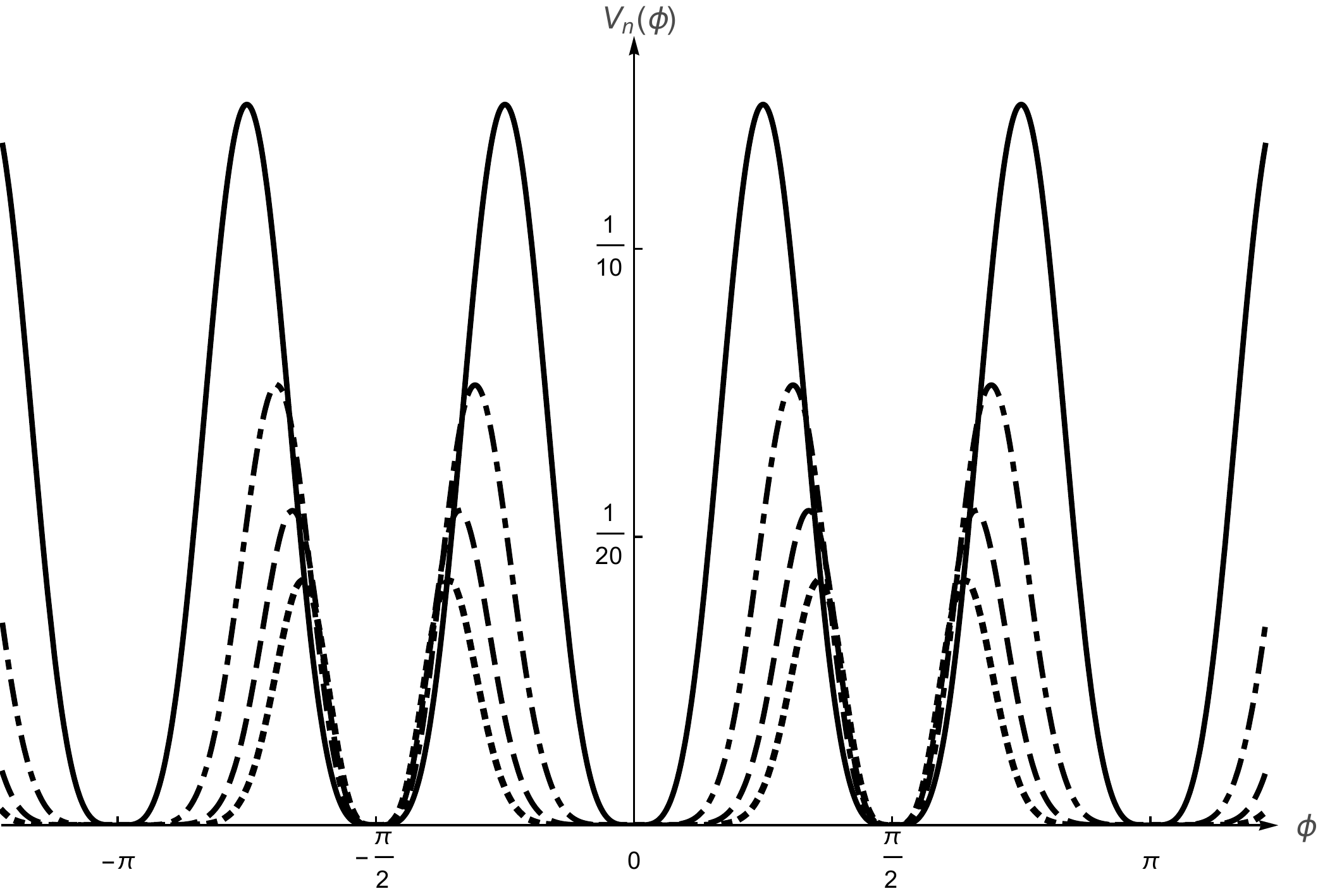}\label{seA}}
{ \includegraphics[width=0.46\linewidth]{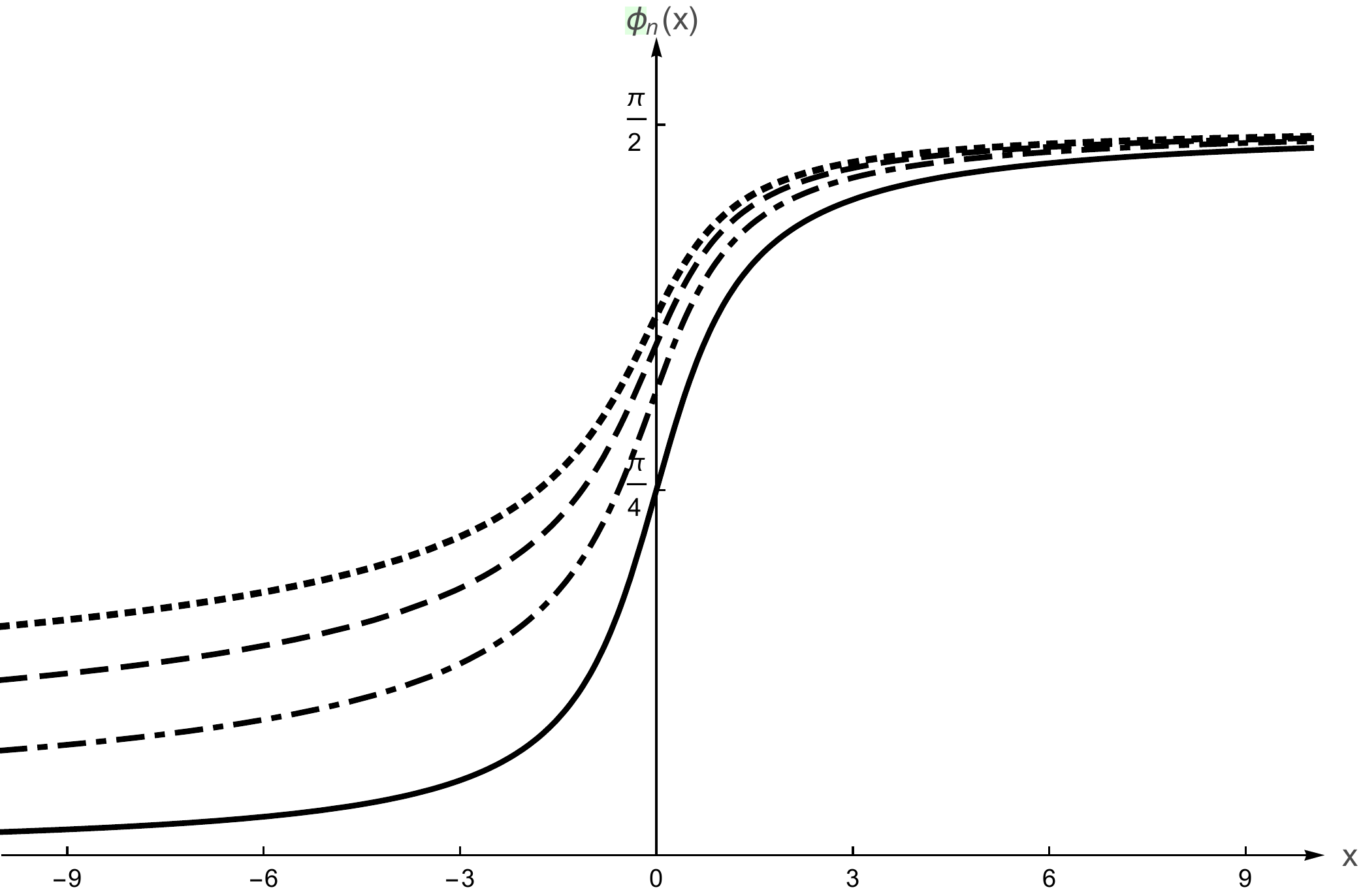}\label{seB}}
{\vspace{0.5cm}}
{\includegraphics[width=0.47\linewidth]{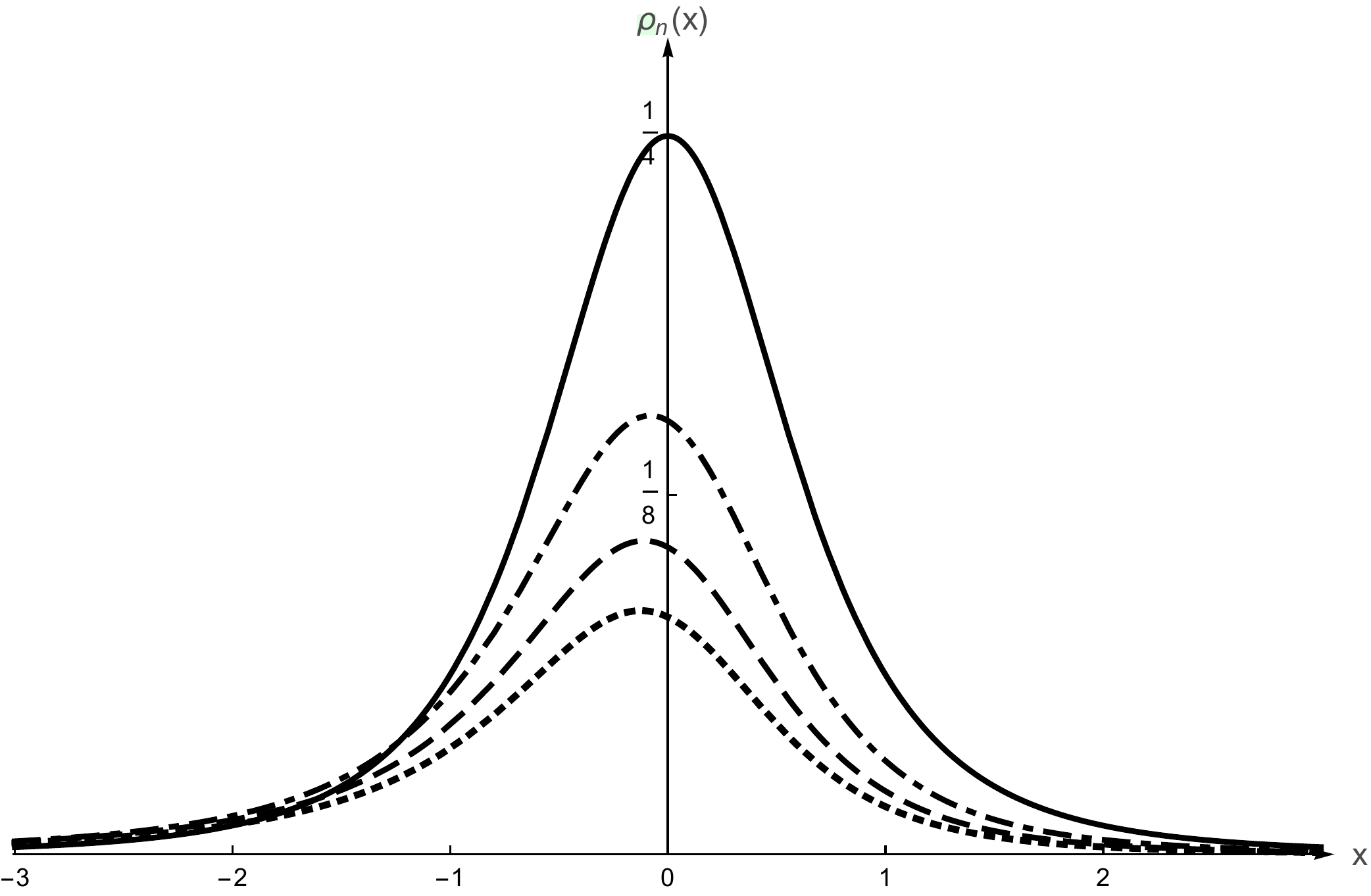}\label{seC}}
{\includegraphics[width=0.47\linewidth]{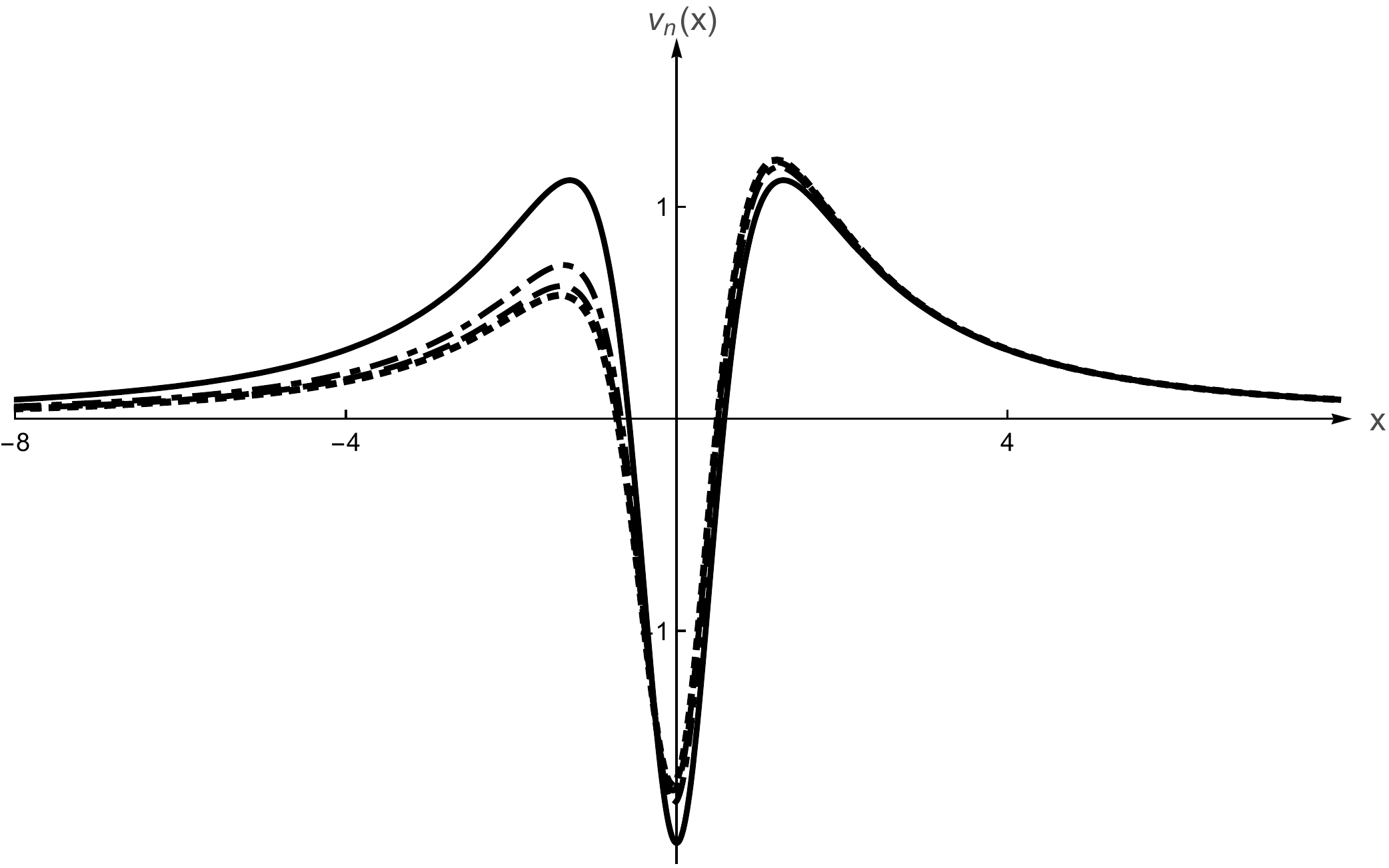}\label{seD}}
\caption{Illustration of the potential (top left), field solution (top right), energy density (bottom left) and stability potential (bottom right) for the model \eqref{pot2}. The solid, dot-dashed, dashed and dotted lines refer to $n=1$, $n=2$, $n=3$ and $n=4$, respectively.\label{fig3} }
\end{figure}

The second model to be investigated in this work is given by the potential
\begin{equation}\label{pot2}
V_n(\phi)=\frac{2}{n^2} \tan^2(\phi)\sin ^{2 n}(\phi) \left(1-\sin ^{2 n}(\phi )\right)^3,
\end{equation}
which is depicted in Fig.~\ref{fig3}. In this system we have reflection symmetry  $\phi\to-\phi$ and periodicity $\phi\to\phi+p\pi$, where $p\in\mathbb{Z}$, which leads us to two infinite sets of degenerate minima.

The minima are given by the set of points $\phi_j=j\pi/2$ for $ j\in\mathbb{Z}$. A straightforward calculation shows that the mass of the scalar field vanishes at any of these minima. Also, there are two subsets of minima, according to the behavior in response to variations of $n$. The behavior of the minima with even and odd $j$ depends on $n$ differently, as it is illustrated in Fig. \ref{fig3}. Thus, the parameter $n$ imposes different power laws in the asymptotic regime of the scalar field for $n\neq 1$, and this implies the existence of kinks with an asymmetry which is different from the one present in the previous model. We note, in particular, that for $n=1$ the model becomes essentially the model \eqref{modelexample} and the two subsets of minima add to become a single set.

The equation of motion derived from the potential \eqref{pot2} is written as follows
\begin{eqnarray}
\nonumber \phi''\!&=&\!\frac{4}{n^2} \tan (\phi) \sin ^{2 n}(\phi) \left(1-\sin ^{2 n}(\phi)\right)^2 \!\times\\
&~&\times\!\left(n+\sec ^2(\phi)-\sin ^{2 n}(\phi) \left(4 n+\sec ^2(\phi)\right)\right)\!.\,
\end{eqnarray}
Again, it is possible to simplify the investigation by introducing an auxiliary function $ W_n(\phi) $. Unfortunately, we could not find it in a closed form for all $n$, but since the potential \eqref{pot2} can be regarded as the square of some function, one can infer that $ W_n(\phi)$ exists. Thus, calculating this function for specific values of $n$ we can write, for instance

\begin{eqnarray}
\nonumber W_{n=1}(\phi)\!\!&=&\!\!\frac{1}{16} (4 \phi -\sin (4 \phi ));\\
\nonumber W_{n=2}(\phi)\!\!&=&\!\!\frac{1}{8}\!\sin^{\!-\!1}\!\!\left(\!\frac{\cos(\phi )}{\sqrt{2}}\!\right)\!+\!\frac{\sqrt{2\!-\!\cos^2 \!(\phi )}}{1536}\Big(233 \cos(\phi)\!+\!\Big.\\
&~&\Big. +51 \cos (3 \phi )-31 \cos (5 \phi )+3 \cos (7 \phi )\Big)\!.\,
\end{eqnarray}

Given the existence of the auxiliary function, one can write the first order equation for the system, which is given by
\begin{equation}
\phi'=\frac{2}{n}\tan(\phi)\sin ^{n}(\phi) \left(1-\sin ^{2 n}(\phi )\right)^{3/2}.
\end{equation}
Solving this equation leads us to the set of solutions
\begin{equation}\label{solu3}
\phi_{n,p}^{\pm}(x)=\pm\arcsin\left(\beta_n^{\pm}(x)\right)+p\pi, ~~p\in \mathbb{Z},
\end{equation}
where, for simplicity, we used
\begin{equation}
\beta_n^{\pm}(x)=\left(\frac{1}{2}\left(1\pm\frac{x}{\sqrt{1+x^2}}\right)\right)^{\frac{1}{2 n}},
\end{equation}
with $\beta_n^{\pm}(0)=2^{-1/2n}$. Note that the signals $ \pm \arcsin(\cdots)$ separates the two types of solutions related to the subsets of topological sectors, while $\beta_n^{\pm}$ identify kink and anti-kink at the specific sector. As the model contains two sets of sectors with degenerate copies, we can concentrate on the ones having $p=0$. Furthermore, the invariance of the model under $\phi\to-\phi $ allows us to restrict our study to the case $\phi^+_{n,0}(x)=\phi_n(x)$ and, for convenience, we work with the kink solution, i.e., we take $\beta_n^+(x)=\beta_n(x)$. The solutions are depicted in Fig.~\ref{fig3}.

Turning attention to the asymptotic behavior of the solution, we find that
$$\phi_n(x)\!\simeq\!\left\{\!\!
\begin{array}{c}
\frac{\pi}{2}-\frac{1}{2\sqrt{n}}\frac{1}{x}\left(1-\frac{1+4n}{24n}\frac{1}{x^2}+\cdots\!\right)\!, \text{if}~x\!\to\!\infty,\\
{}\\
\frac{1}{\left(2|x|\right)^{1\!/\!n}}\!\left(1+\frac{1}{6}\frac{1}{\left(2|x|\right)^{2\!/\!n}}+\!\cdots\!-\frac{3}{4x^2}\left(\alpha_1+\right.\right.\\
{}\\
\left.\left.+\alpha_2\frac{1}{\left(2|x|\right)^{2/n}}+\cdots\right)+\cdots\right)\!, \text{if}~x\to-\infty,
\end{array}
\right.$$\\
with $\alpha_1={1}/{(2n+1)}$ and $\alpha_2={1}/{(2(2n+3))}$. For $ n = 1 $ we have solutions with symmetric behavior around $x = 0$, as expected. For $n\neq1$ both kink tails decay as a power law, as it is shown in the above expression. This family of models presents another possibility to get kinks with polynomial tails, which could be further explored concerning their collective physical behavior \cite{SM,QG,QI}.

The energy density of the solution is
\begin{eqnarray}\label{ro2}
\rho_n(x)\!=\!\frac{1}{4n^2}\frac{1}{1+x^2}\left(1-\frac{x}{\sqrt{1+x^2}}\right)^2\frac{\beta_n^2(x)}{1-\beta_n^2(x)}
\end{eqnarray}
which is depicted in Fig.~\ref{fig3}. Note that variations in $ n $ does affect the way the energy density dips into the coordinate space, reinforcing the asymmetry of $\rho_n(x)$.  Moreover,  asymptotically we find
$$\rho_n(x)\simeq\left\{
\begin{array}{c}
\frac{1}{4n}\frac{1}{x^4}\left(1-\frac{1+4n}{4n}\frac{1}{x^2}+\cdots\right)\!, \text{if}~x\to\infty,\\
{}\\
\!\!\!\!\!\!\!\!\frac{1}{n^2}\frac{1}{x^2}\left(\frac{1}{\left(2|x|\right)^{2/n}}+\frac{1}{\left(2|x|\right)^{4/n}}+\cdots\right)\times\\
{}\\
~~~~\times\left(1-\frac{3}{2x^2}+\cdots \right)\!, ~~~~\text{if}~x\to-\infty.
\end{array}
\right.$$\\
Here we observe that while in the $x\to\infty$ regime the energy density decays as $\sim 1/x^4$, in the opposite regime we have responses to the variations of the parameter of the model with a dominant term given by $\sim 1/|x|^{2+2/n}$. The stability potential of this model is written as follows 
\begin{eqnarray}
\nonumber v_n(x)&=&\frac{1 }{\left(1+x^2\right)^2 }\left( \left(7 x^2+5 x\sqrt{1+x^2} -1\right)-\right.\\
\nonumber&~&-\frac{3}{2n} \left(1-x \left(x-\sqrt{1+x^2}\right)\right) \frac{1}{\left(1-\beta_n^2(x)\right)}+\\
\nonumber&~&\left.+\frac{1}{4n^2}\left(x-\sqrt{1+x^2}\right)^2 \frac{\left(1+2\beta_n^2(x)\right)}{\left(1-\beta_n^2(x)\right)^2}\right)
\end{eqnarray}
and is also depicted in Fig.~\ref{fig3}. We find an asymmetric volcano potential for $n\neq1$.  As it approaches zero when $ x\to\pm\infty $, we have no bound states other than the zero mode, which can be written in the form
\begin{equation}
\eta_0(x)=\frac{1}{2n}\frac{\sqrt{1+x^2}-x}{1+x^2}\frac{\beta_n(x)}{\sqrt{1-\beta_n^2(x)}}.
\end{equation}

\section{Ending comments} 

In this work we investigated models that support analytical solutions which are derived from potentials having massless minima. Models presenting this characteristic have properties that are distinct from the majority of analytical models presented so far, which present exponential decay. The solutions investigated previously in Refs.~\cite{riazi,gmo} are symmetric and have the decay laws of the form $ \simeq 1/|x| $ and $\simeq 1/x^2$. In this sense, the models investigated in the current work are more general: they present analytical solutions that are asymmetric, having power law tails which depend on the parameter that control how the scalar field self-interacts. The localized structures may then be used to model collective behavior of Bose-Einstein condensates and Rydberg atoms, among other possibilities \cite{SM,QG,QI}.

We studied models with two distinct asymmetries. In the first case, the models considered have topological sectors that connect distinct minima, one having zero mass and the other non-vanishing mass. In the second case, both minima have vanishing masses, but we could control the order of the first nonzero derivative at the minima, so the asymmetry arises due to the differences between the power law at the minima. In both cases, we found stability potentials that engender only one bound state, the zero mode, in a way consistent with the fact that the stability potential tends asymptotically to the masses at the corresponding minima. 

The above results suggest new investigations, one of them being related to the model I, which is somehow similar to the $\phi^6$ model, so it should also be examined as in Ref.~\cite{dorey}, to see how the kink and anti-kink collisions respond to this new scenario. Here we recall that in \cite{dorey} the authors considered the standard $\phi^6$ model, which supports localized structure that decays asymmetrically, but engendering short range tails with exponential behavior. In the new model defined by the potential \eqref{pot1}, the localized structure also decays asymmetrically, but now it may have long range tail that will certainly contributes differently. Another possibility concerns the motion of kinks in the model with three degenerate minima in the presence of radiation, to see how the radiation works to destabilize the topological structure, in a way similar to the case implemented in the recent study \cite{TR}. 

We can also ask about the collision of kinks and anti-kinks with polynomial tails, an issue that can be implemented as in Ref.~\cite{gani6}. Since the topological structures here investigated simulate long range interactions, further investigations are welcome, mainly in the directions concerning issues related to their collective behavior \cite{SM}, and to applications to the physics of quantum gases \cite{QG} and quantum information \cite{QI}. The long range interactions that we found in this work decay polynomially and will certainly bring novelties, concerning the investigation of the scattering of kinks and anti-kinks. This is currently under consideration in \cite{bbg}, following the lines of Ref.~\cite{bbg0}, and we hope to describe how the parameter $n$ contributes to control the kink/anti-kink collisions.

The authors would like to thank the Brazilian agencies CAPES and CNPq for financial support.


\end{document}